\documentclass [aps,twocolumn,superscriptaddress,showpacs,preprint]{revtex4}
\usepackage{graphicx}
\usepackage{amsmath,amsfonts,amsfonts}

\begin{document}
\title {Pressure-induced spin transition of Fe$^{2+}$ ion in ferropericlase from Dynamical Mean-Field Theory }

\author {N.~A.~Skorikov}
\affiliation {M.N. Miheev Institute of Metal Physics of Ural Branch of Russian Academy of Sciences,
620137 Yekaterinburg, Russia}

\author {A.~O.~Shorikov}
\affiliation {M.N. Miheev Institute of Metal Physics of Ural Branch of Russian Academy of Sciences,
620137 Yekaterinburg, Russia}
\affiliation {Ural Federal University, 
620002 Yekaterinburg, Russia}

\author {S.~L.~Skornyakov}
\affiliation {M.N. Miheev Institute of Metal Physics of Ural Branch of Russian Academy of Sciences,
620137 Yekaterinburg, Russia}
\affiliation {Ural Federal University, 
620002 Yekaterinburg, Russia}

\author {M.~A.~Korotin}
\affiliation {M.N. Miheev Institute of Metal Physics of Ural Branch of Russian Academy of Sciences,
620137 Yekaterinburg, Russia}

\begin{abstract}
The results of magnetic and spectral properties calculation for paramagnetic phase of ferropericlase (Fe$_{1/4}$Mg$_{3/4}$)O at ambient and high pressures are reported. Calculations were performed by combined Local Density Approximation + Dynamical Mean-Field Theory method (LDA+DMFT). At ambient pressure calculation gave (Fe$_{1/4}$Mg$_{3/4}$)O as insulator with Fe 3$d$-shell in high-spin state.
Experimentally observed high-spin  to low-spin transition of Fe$^{2+}$ ion at high pressure is successfully reproduced in calculations.
Our results indicate the absence of metal-insulator transition up to the pressure 107~GPa.
\end{abstract}

\maketitle{}

\section{Introduction}
Ferropericlase (MgFe)O forms a solid solution between periclase (MgO) and wustite (Fe$_{1-x}$O) and crystallizes in cubic rock-salt structure. 
Ferropericlase (Mg,Fe)O is a major lower mantle mineral, and its high-pressure and high-temperature properties are of great importance for the Earth science. 
Wustite is a Mott insulator and periclase is a wide band gap insulator. At the normal conditions (MgFe)O is a paramagnetic insulator, which the Neel temperature depends of Fe content.
In ferropericlase oxidation state of Fe is Fe$^{2+}$ (d$^{6}$ configuration). The d$^{6}$ configuration may be realized in two  states with different spin: the high-spin (HS) state (S=2) and the low-spin (LS) state (S=0). 

If the Hund's exchange energy prevails the crystal field splitting then the HS state is realized, otherwise ions are in the LS state. The spin state can change depending on the ratio of the crystal field and exchange energy (for example, under pressure). Spin-state changes of Fe$^{2+}$ ion in ferropericlase will influence many of its properties, including bulk density, elasticity and electrical conductivity. For these reasons, a knowledge of the pressure dependence and characteristics of the spin crossover transition in ferropericlase is necessary for understanding the constitution of the Earth’s lower mantle. 

The spin crossover of Fe$^{2+}$ ion in (MgFe)O under pressure was intensively studied both experimentally and theoretically \cite{Sherman1991, Sherman1995, Wentzcovitch2009, Koci2007, Sun2008, Sturhahn2005, Badro2003, Lin2005, pnas.1304827110, Gavriliuk2006, Fei2007, Lin2007, MIT-exp}. At a pressure near the spin crossover ions in the HS and LS states may exist simultaneously due to temperature fluctuations. Such a mixed state leads to an extended transition region width of the order of 20~GPa  \cite{pnas.1304827110}. According to experimental data \cite{pnas.1304827110} pressure of start and end of the spin crossover  as well as its width are dependent on temperature. In various articles one can find different experimental estimates of transition pressure. For room temperature the spin crossover may start as low as 35 GPa \cite{Fei2007} and end as high as 75~GPa \cite{Badro2003}. At temperature $\approx$500~K the HS-LS transition occurs in the pressure range 40-70~GPa \cite{pnas.1304827110}. 

Another effect of high pressure is the metal-insulator transition (MIT). The MIT under pressure is observed, for example, in FeO, MnO, Fe$_{2}$O$_{3}$ \cite{feo, mno, fe2o3}. Metallization of (MgFe)O with pressure has been predicted by theory \cite{MIT-theory1, MIT-theory2} but not observed experimentally 
at least up to 140~GPa and 2100~K \cite{Lin2007, MIT-exp}. Both the HS-LS and the MIT in MnO and Fe$_{2}$O$_{3}$ with pressure can be successfully described using the LDA+DMFT calculation scheme \cite{mno, fe2o3}. Presence of the MIT and absence of the HS-LS with pressure in FeO is described in the LDA+DMFT calculations also \cite{feo}. In the present paper we  show that the LDA+DMFT method reproduces the HS-LS transition for (MgFe)O with pressure. At the same time the HS-LS transition in (MgFe)O in contrast to MnO and Fe$_{2}$O$_{3}$ is not accompanied by MIT. 

\section{Method}
The LDA+DMFT calculation scheme \cite{LDA+DMFT} is constructed in the following way: first, a Hamiltonian $\hat H_{LDA}$ is produced using converged LDA results for the system under investigation, then the many-body Hamiltonian is set up, and finally the corresponding self-consistent DMFT equations are solved.

The basis of calculation is the supercell containing 8 formula units of MgO in which two Mg atoms were replaced by two Fe atoms. The positions of the iron atoms were chosen so that the impurity atoms are maximally spaced from each other. To calculate the electronic structure of ferropericlase at ambient pressure the experimentally determined parameters of the crystal structure of  (Fe$_{0.24}$Mg$_{0.76}$)O were used \cite{Jacobsen2002}. The application of hydrostatic pressure simulated by supercell volume reduction with 5\% steps. For each volume parameter of the Coulomb repulsion U was calculated \cite{U-calc} and a low-dimensional Hamiltonian in the basis of Wannier functions using the projection procedure \cite{wf_proj} was recorded. Hamiltonian includes the Fe-$3d$ and O-$2p$ states and has a dimension 34$\times$34. The obtained values ​​of the parameter of the Coulomb repulsion U vary in the range from U=4.9~eV for the equilibrium volume cell to U=4.24~eV for the smallest considered cell, the volume of which is reduced to 70\% relative to the original. The exchange interaction parameter J is not changed at varying the volume of the unit cell, it was selected J=0.95~eV. The LDA calculations were performed using  the TB-LMTO-ASA (Tight Binding-Linearized Muffin-Tin Orbitals-Atomic Sphere Approximation) code \cite{tb-lmto}. 

The relation between the volume of the unit cell and pressure was made according to data \cite{Lin2005}. The calculations  presented below  have been done for crystal volumes corresponding to  values of pressure up to 107~GPa. 
        
The many-body Hamiltonian to be solved by the DMFT has the form
\begin{equation}
\hat H= \hat H_{LDA}- \hat H_{dc}+\frac{1}{2}\sum_{i,\alpha,\beta,\sigma,\sigma^{\prime}}
U^{\sigma\sigma^{\prime}}_{\alpha\beta}\hat n^{d}_{i\alpha\sigma}\hat n^{d}_{i\beta\sigma^{\prime}},
\label{eq:ham}
\end{equation}
where $U^{\sigma\sigma^{\prime}}_{\alpha\beta}$ is the Coulomb interaction matrix, 
$\hat n^d_{i\alpha\sigma}$ is the occupation number operator 
for the $d$ electrons with orbitals $\alpha$ or $\beta$ and spin indices $\sigma$ 
or $\sigma^{\prime}$ on the $i$-th site. 
The term $\hat H_{dc}$ stands for the {\it d}-{\it d} interaction 
already accounted for in the LDA, so called double-counting correction. 
In the present calculation the double-counting was chosen in 
the following form $\hat H_{dc}=\bar{U}(n_{\rm dmft}-\frac{1}{2})\hat{I}$.
Here $n_{\rm dmft}$ is the self-consistent total number of {\it d} electrons 
obtained within the LDA+DMFT, $\bar{U}$ is the average Coulomb parameter for 
the {\it d} shell. 

The elements of $U_{\alpha\beta}^{\sigma\sigma'}$ matrix
are parameterized by $U$ and $J_H$ according to procedure described in
\cite{LichtAnisZaanen}. 
The effective  impurity problem for the DMFT was solved by the hybridization  expansion Continuous-Time Quantum Monte-Carlo method (CT-QMC) \cite{CTQMC}.  Calculations for all volumes were performed in the paramagnetic state at the inverse temperature $\beta=1/kT$=20 eV$^{-1}$ corresponding to 580~K. Spectral functions on real energies  were calculated by Maximum Entropy Method (MEM)\cite{mem}.

\section{Results and discussion}
\begin {figure}[htb!]
\centering
\includegraphics[width=0.90\linewidth]{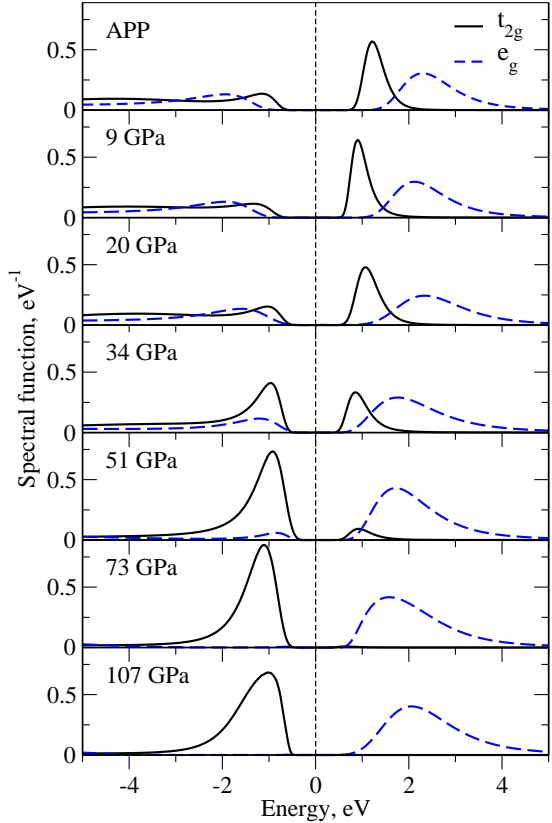}
\caption {(Color online) Spectral function of Fe-d states vs. pressure obtained in the LDA+DMFT (CT-QMC) calculations.}
\label {fig:dos}
\end{figure}

In octahedral coordination the Fe {\it d} band is split by crystal field in triply degenerated $t_{2g}$ and doubly degenerated $e_g$ subbands. Spectral functions for all pressure values calculated by MEM from CT-QMC calculations are presented in Fig.~\ref{fig:dos}. The spectral function for ambient pressure phase (APP) shows well defined insulating behavior. The energy gap for $t_{2g}$ is smaller than that for $e_{g}$ states. The top of valence band and bottom of conduction band are formed by $t_{2g}$ states. The value of band gap is $\Delta$E$_{gap}$=1.13~eV.

\begin {figure}[htb!]
\vspace{5mm}
\includegraphics [width=0.9\linewidth]{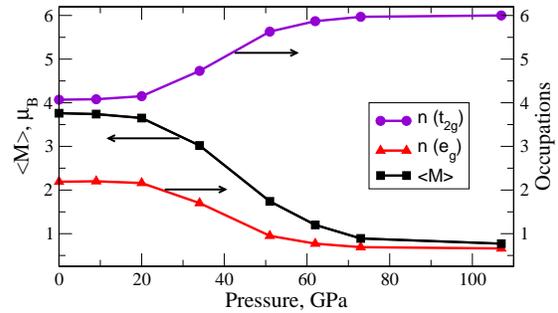}
\caption {(Color online) Magnetic moments (black squares) and occupancies of $t_{2g}$ (magenta circles) and $e_g$ (red triangles) shells vs. pressure obtained in the LDA+DMFT (CT-QMC) calculations. }
\label {fig:occ}
\end {figure}

For APP the paramagnetic insulator state was obtained, the average value of  local magnetic moment $\sqrt{<\mu_z^2>}$ is 3.76$\mu_B$, states with $t_{2g}$ and $e_{g}$ symmetry are partially filled. The  occupation numbers  for Fe d orbitals are n($e_{g}$)=2.16 and n($t_{2g}$)=4.14 (Fig.~\ref{fig:occ}). This agrees well with high-spin state of Fe$^{2+}$ ion in cubic crystal field.

With the pressure growth, the population of orbital with $e_{g}$-symmetry becomes energetically unfavorable and the filling of $t_{2g}$-orbitals starts. At pressure 34~GPa the first noticeable changes arise: the occupancy of $e_{g}$ orbitals and the average value of  local magnetic moment decrease, occupancy of $t_{2g}$ orbitals increases (Fig.~\ref{fig:occ}). The pronounced maximum of the  $t_{2g}$ spectral function is formed in valence band (Fig.~\ref{fig:dos}).

At pressure 73~GPa the redistribution of electrons between the $e_{g}$ and $t_{2g}$ orbitals is completed and all 6 electrons of Fe$^{2+}$ ion are localized at orbitals of $t_{2g}$-symmetry. At this pressure the average value of local magnetic moment $\sqrt{<\mu_z^2>}$ becomes 0.89$\mu_B$ (Fig.~\ref{fig:occ}). The valence band is formed by the $t_{2g}$ states, the conduction band is formed by the $e_{g}$ states (Fig.~\ref{fig:dos}). That is completely different mechanism of the band gap formation for the LS state in comparison with that for HS state.
At the end of the transition the value of energy gap is reduced to $\Delta$E$_{gap}$=0.65~eV, the metal-insulator transition does not occur up to a maximum considered pressure 107~GPa. 

The crystal structure of ferropericlase is completely analogous to the structure of  FeO. In both cases Fe$^{2+}$ ion is in an octahedral environment of oxygen atoms, in both cases the top of the valence band and the bottom of the conduction band is formed by the d-Fe states. In FeO metallic spectral function is observed only for $t_{2g}$ orbitals \cite{feo}. The absence of MIT in ferropericlase in contrast to FeO could be explained as follows. Unlike FeO, in (Fe$_{1/4}$Mg$_{3/4}$)O there are no direct overlap of the d-Fe orbitals, so the Fe-$t_{2g}$ bands are significantly narrower than in FeO (in the LDA width at half maximum is less then 0.1~eV, appropriate figure is not shown). Under pressure Fe-$t_{2g}$ bands becomes wider, but the pressure 107~GPa is not enough for the metallization of the spectrum.
 
%\begin {figure}[htb!]
%\centering
%\includegraphics[width=0.90\linewidth]{fig3.eps}
%\caption {(Color online) LDA densities of states (DOS) ($e_g$ thin solid lines, $t_2g$ bold solid lines). Upper panel correspond to APP conditions, bottom panel -- high pressure condition.}
%\label {lda:dos}
%\end{figure}

%According to experimental measurements the Fe$^{2+}$ ions undergo a high-spin to low-spin transition in (Fe$_{1/4}$Mg$_{3/4}$)O under pressure 51-73 GPa, which correspond to a decrease in the unit cell volume from 80\% to 75\% relative to the equilibrium \cite{Lin2005}.

As mentioned in the Introduction, there are different experimental evaluations of the pressure at which the magnetic spin transition occurs.  Our theoretical estimate of the transition pressure interval 34-73~GPa is in reasonable agreement with experimental results \cite{Lin2005,pnas.1304827110}. 
In our calculation the supercell approximation to construct (Fe$_{1/4}$Mg$_{3/4}$)O was used, which makes artificial spatial ordering of iron atoms. Promising continuation of the present study seems the LDA+DMFT+CPA method, which allows a natural way to take into account disorder in the sublattice of Fe-Mg. 

\section{Conclusion}
We have performed the LDA+DMFT calculation for (Fe$_{1/4}$Mg$_{3/4}$)O at  580~K and values of pressure from the ambient one up to 107~GPa. The HS to LS transition of the Fe$^{2+}$ ion starts at the pressure 34~GPa and at the pressure 73~GPa transition is completed. At the  pressure 107~GPa the value of band gap decreases till 0.7 eV, but in contrast to FeO the metal-insulator transition does not occur.

\section*{Acknowledgments}
This study was supported by the grant of the Russian Scientific Foundation (project no. 14-22-00004).

\end{document}